\documentclass[10pt,journal,comsoc]{IEEEtran}

\usepackage{amsmath,amssymb,amsthm}
\usepackage{graphicx}
\usepackage{subfigure}
\usepackage{curves}
\usepackage{amsfonts}
\usepackage{epsfig}
\usepackage{stfloats}
\usepackage{cite}
\usepackage{algorithm}
\usepackage{algorithmic}
\usepackage{epstopdf}
\usepackage{multirow}
\usepackage{array}
\usepackage{rotating}
\usepackage{comment}

\begin{document}

\title{Polar Coding and Sparse Spreading for Massive Unsourced Random Access}
\author{Mengfan~Zheng, Yongpeng~Wu and Wenjun Zhang
	

\thanks{M. Zheng is with the Department of Electrical and Electronic Engineering at Imperial College London, United Kingdom. Email: m.zheng@imperial.ac.uk.}	
\thanks{Y. Wu (corresponding author) and W. Zhang are with the Department of Electronic Engineering at Shanghai Jiao Tong University, Shanghai, China. Emails: \{yongpeng.wu, zhangwenjun\}@sjtu.edu.cn.}
\thanks{The work of Y. Wu is supported in part by the National Key R\&D Program of China under Grant 2018YFB1801102,  JiangXi Key R\&D Program under Grant 20181ACE50028, National Science Foundation (NSFC) under Grant 61701301, the open research project of State Key Laboratory of Integrated Services Networks (Xidian University) under Grant ISN20-03, and Shanghai Key Laboratory of Digital Media Processing and Transmission (STCSM18DZ2270700). The work of W. Zhang is supported by  Shanghai Key Laboratory of Digital Media Processing and Transmission (STCSM18DZ2270700). }
}

\maketitle

\begin{abstract}
  In this paper, we propose a new polar coding scheme for the unsourced, uncoordinated Gaussian random access channel. Our scheme is based on sparse spreading, treat interference as noise and successive interference cancellation (SIC). On the transmitters side, each user randomly picks a code-length and a transmit power from multiple choices according to some probability distribution to encode its message, and an interleaver to spread its encoded codeword bits across the entire transmission block. The encoding configuration of each user is transmitted by compressive sensing, similar to some previous works. On the receiver side, after recovering the encoding configurations of all users, it applies single-user polar decoding and SIC to recover the message list. Numerical results show that our scheme outperforms all previous schemes for active user number $K_a\geq 250$, and provides competitive performance for $K_a\leq 225$. Moreover, our scheme has much lower complexity compared to other schemes as we only use single-user polar coding. 
  
\end{abstract}


\section{Introduction}
    
  The study of massive random-access has drawn increasing attention recently as massive machine-type communications (mMTC) is one of the key application scenarios in 5G and beyond \cite{wu2020massive}. Polyanskiy introduced the unsourced random access channel (RAC) in \cite{Polyanskiy2017RA} for modelling the problem of providing multi-access for a massive number of infrequent and uncoordinated users. The major deviations of this model from traditional multiple access channel (MAC) are as follows. Since the user number is very large but only a small fraction are active at a time, and the payloads of mMTC devices are usually very short, it is inefficient to do user identification in each transmission. Therefore, the decoder’s job is to only produce a list of transmitted messages, not to identify from whom they are sent. Correspondingly, all users use the same codebook to encode their messages. To measure the average fraction of correctly decoded messages, the error performance is evaluated by per-user probability of error (PUPE) instead of global error probability. In \cite{Polyanskiy2017RA}, an achievability bound of this channel was derived and the asymptotic problem  (finite blocklength effect) was also studied. 
  
  Since the publication of \cite{Polyanskiy2017RA}, there have been several attempts to design low-complexity schemes to approach that bound \cite{Ordentlich2017RA,Vem2017RA,amalladinne2018coupled,amalladinne2018coded,fengler2019sparcs,calderbank2018chirrup,pradhan2019joint,marshakov2019polar,pradhan2019polar}. The ideas include $T$-fold ALOHA \cite{Ordentlich2017RA,Vem2017RA,calderbank2018chirrup,marshakov2019polar} (which divides a transmission block into sub-blocks and allows up to $T$ collision users in each sub-block), splitting the payloads into smaller pieces and recovering them by a tree code\cite{amalladinne2018coupled,amalladinne2018coded,fengler2019sparcs}, sparsifying codeword bits among the whole block to reduce interference \cite{pradhan2019joint}, and mitigating multi-user interference by random spreading \cite{pradhan2019polar}. 
  
  One lesson learned from information theory is that to achieve lower energy-per-bit, one needs to use longer code with lower rate. This might explain why most of the previous schemes (except two polar coding schemes of \cite{marshakov2019polar,pradhan2019polar}) perform quite poorly even when the active user number $K_a$ is small, because they all adopt short or moderate code-lengths so as to reduce interference. The random spreading scheme of \cite{pradhan2019polar}, however, uses spreading sequences to disperse codeword bits into the whole block. This scheme effectively creates longer codes and gives the best performance so far when $K_a\leq 225$. Nevertheless, the performance of this scheme deteriorates drastically as $K_a$ increases and becomes uncompetitive when $K_a\geq 250$.  To the authors' understanding, this could partly be due to the code-length being too long. Reference \cite{pradhan2019polar} deals with multi-user interference by treat interference as noise with successive interference cancellation (TIN-SIC). Since each user occupies all channel uses, multi-user interference becomes harder and harder to handle as the active user number increases.
  
  In this paper, we propose a new polar coding scheme for the Gaussian RAC based on sparse spreading and TIN-SIC. We do not use slotted ALOHA as it will limit the maximum code-length that can be used. Instead, we adopt the approach of \cite{pradhan2019joint} to sparsify encoded bits using interleavers. In TIN-SIC schemes, codeword can be modulated using several power levels. In our scheme, we further introduce variable-length encoding to offer more flexibility of SIC, i.e., messages can be encoded into codewords with different code-lengths for each power level. Each active user randomly selects its encoding, modulation and interleaving method according to some probability distribution designed to allow the receiver to perform TIN-SIC. On the receiver side, it first decodes the longest codewords among those with the highest power $P_s$, treating all other users' signals as noise. Then it subtracts the decoded codewords from the received signal and decodes the second longest codewords with power $P_s$. After decoding all codewords with power $P_s$, it then moves to codewords with the second highest power, and so on and so forth until all codewords are decoded. Another benefit of using various code-lengths with the same power is that it simplifies the estimation of multi-user interference. As we know, a binomial distribution can be well approximated by a Gaussian distribution when the sample number is sufficiently large. We show that the multi-user interference in our scheme can also be well approximated by Gaussian noise, which helps simplify our code design and decoding. 
  
  We show by numerical experiments that our scheme outperforms all existing schemes for $K_a\geq 250$, and is only inferior to the scheme of \cite{pradhan2019polar} for $K_a\leq 225$. Moreover, since we only use single-user polar codes and SIC in our scheme, the per-user decoding complexity is just $O(LN\log N)$, where $N$ is the code-length and $L$ is the decoding list size. Besides, polar codes in our scheme can be efficiently constructed using existing low-complexity methods. This is a great advantage over previous schemes.

\section{Preliminaries}
\label{S:Preliminaries}

 \subsection{System Model}
 
 In the considered channel model, there are $K_{tot}$ users in total, out of which only $K_a$ are active at a time, where $K_a\ll K_{tot}$. Each active user wishes to send $B$ information bits to an access point over $N$ channel uses. Then the input-output relation can be written as
 \begin{equation}
 \mathbf{y}=\sum_{i=1}^{K_{tot}}s_i \mathbf{x}_i(\mathbf{u}_i)+\mathbf{z},\label{ChannelModel}
 \end{equation}
 where $s_i\in \{0,1\}$ indicates whether user $i$ is active or not, $\mathbf{u}_i\in\{0,1\}^B$ and $\mathbf{x}_i(\mathbf{u}_i)$ respectively are the message bits and codeword of user $i$, $\mathbf{y}$ is the received signal by the access point, and $\mathbf{z}\sim \mathcal{N}(0,\mathrm{\mathbf{I}}_N)$ is white Gaussian noise. Users' transmissions are under an average power constraint, i.e., $\mathbb{E}(\|\mathbf{x}_i\|^2)\leq NP$. The energy-per-bit of the system is defined as 
 $$E_b/N_0\triangleq \frac{NP}{2B}.$$ 
 The receiver's task id to produce a list of transmitted messages $\mathcal{L}(\mathbf{y})$, without having to identify the sources of them. The per-user probability of error (PUPE) of the system is given by 	
 \begin{equation}
 	P_e=\max_{\sum s_i =K_a} \frac{1}{K_a}\sum_{i=1}^{K_{tot}}s_i \Pr(\mathbf{u}_i\notin \mathcal{L}(\mathbf{y})).
 \end{equation}
 For fixed $N$, $B$, $K_a$ and a target error probability $\epsilon$, the objective is to design a low-complexity scheme that satisfies $P_e\leq \epsilon$ at the lowest $E_b/N_0$ possible.

  \subsection{Polar Codes}
  
  The generator matrix of polar codes can be written as
  \begin{equation}
  \mathbf{G}_N=\mathbf{B}_N \textbf{F}^{\otimes n},
  \end{equation}
  where $\otimes$ is the Kronecker power, $\mathbf{B}_N$ is the bit-reversal matrix \cite{arikan2009channel} and $\textbf{F}=
  \begin{bmatrix}
  1 & 0 \\
  1 & 1
  \end{bmatrix}$. The encoding procedure of polar codes is then $$\mathbf{x}=\mathbf{u}\mathbf{G}_N,$$ 
  where $\mathbf{u}=\{u_1,...,u_N\}$ is the uncoded bits and $\mathbf{x}=\{x_1,...,x_N\}$ is the encoded codeword. To construct a polar code, we need to partition $\mathbf{u}$ into an \textit{information bit set} $\mathcal{I}$ and a \textit{frozen bit set} $\mathcal{F}=\mathcal{I}^C$, where $\mathcal{I}^C$ denotes the complement set of $\mathcal{I}$. For symmetric channels, the frozen bits can be simply assigned to 0 \cite{arikan2009channel}. Then the encoding can be written as
  \begin{equation}
  \mathbf{x}=\mathbf{u}^{\mathcal{I}}\mathbf{G}_N^{\mathcal{I}}, \label{PolarEnc}
  \end{equation}
  where $\mathbf{u}^{\mathcal{I}}=\{u_i:i\in \mathcal{I}\}$ and $\mathbf{G}_N^{\mathcal{I}}$ consists of the rows of $\mathbf{G}_N$ indexed by $\mathcal{I}$.
  
  Upon receiving $\mathbf{y}$, the receiver can use a successive cancellation (SC) decoder to decode:
  \begin{equation}
  \bar{u}_{i}=
  \begin{cases}
  u_i,&\text{ if } i\in \mathcal{F}\\
  \arg\max_{u\in\{0,1\}}P_{U_{i}|\mathbf{Y}, U_{1:{i-1}}}(u|\mathbf{y},u_{1:{i-1}}),&
  \text{ if } i\in \mathcal{I}
  \end{cases}.
  \end{equation}
  To improve error performance, we can use the cyclic redundancy check (CRC)-aided successive cancellation list (CA-SCL) decoding \cite{tal2015list}. The idea of CA-SCL is to retain up to $L$ most probable paths during the SC decoding process and use CRC to eliminate wrong paths.

  \subsection{Sparse Spreading and Interference Approximation}
  
  We adopt the sparse spreading scheme introduced in \cite{pradhan2019joint} to reduce multi-user interference. Given a codeword $\mathbf{c}$ of length $N_c<N$, we first zero-pad it to a length $N$ vector $\mathbf{c}'$,
  $$\mathbf{c}'=[\mathbf{c},0,...,0].$$
  Then we apply a random interleaver $\pi$ on it and obtain the final codeword $\mathbf{x}$: $$\mathbf{x}=\pi(\mathbf{c}').$$
  In this way, the effective multi-user interference at bit level is reduced. Suppose there are $K$ users adopting length-$N_c$ codes and using the same modulation power (normalized to 1 here). The probability that there are $k$ user occupying any specific bit position is given by
  \begin{align}
  P_c(k)=\binom{K}{k}\Big{(}\frac{N_c}{N}\Big{)}^k\Big{(}1-\frac{N_c}{N}\Big{)}^{K-k}.
  \end{align}
  Assume that the BPSK-modulated codeword bits are uniformly distributed over $\{-1,1\}$ (which can be easily satisfied by polar codes). Then the probability distribution of the sum of all users' signals at a specific bit position is given by
  \begin{align}
  P_{int}(m)=\sum_{i=0}^{\lfloor(K-|m|)/2\rfloor}P_c(|m|+2i)\binom{|m|+2i}{i}\big{(}\frac{1}{2}\big{)}^k, \label{ProbInter}
  \end{align}
  where $m\in\{-K,-K+1,...,K-1,K\}$. The reason behind (\ref{ProbInter}) is that $m$ equals the number of users sending $+1$ minus that of users sending $-1$ at the considered bit position. Thus, to get an interference of $m$, the total number of users occupying that bit position must be of the form $|m|+2i$ for some $0\leq i\leq \lfloor(K-|m|)/2\rfloor$, and there are exactly $|m|+i$ or $i$ users sending $+1$ or $-1$, depending on the sign of $m$.
  
\begin{figure}[tb]
  	\centering
  	\includegraphics[width=9cm]{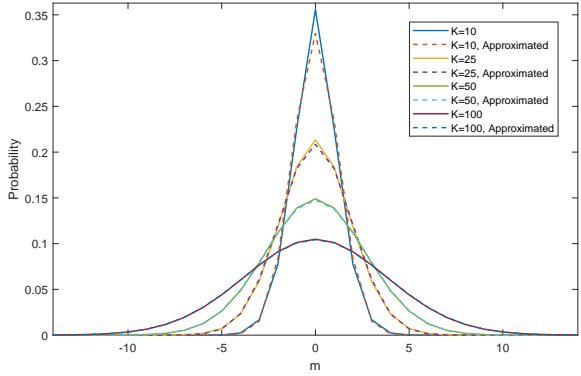}
  	\caption{Distribution of multi-user interference.} \label{fig:muinterference}
  \end{figure}

  It is known that binomial distribution can be approximated by Gaussian distribution if the sample number is large enough. For the distribution of (\ref{ProbInter}), we compare it with Gaussian distribution of zero mean and variance $KN_c/N$ numerically and find that they are very close, even when $K$ is small. An example is shown in Fig. \ref{fig:muinterference}, in which we set $N_c=4096$, $N=28000$, and $K\in\{10,25,50,100\}$. This can simplify our code construction and decoding since we can simply treat interference as Gaussian noise in the TIN phase of our scheme.
  
  \section{Details of the Proposed Scheme}
  \label{S:ProSch}

  \subsection{Codebook and Modulation}
  \label{S:codebook}
  
  In our scheme, codewords are modulated with BPSK of (possibly) multiple power levels, $P_1, P_2,..., P_s$, where $s\geq 1$ and $P_1<P_2<...<P_s$. Messages can be encoded into codewords of various lengths. For codes that are modulated with power $P_j$, denote the candidate code-lengths by $N_j^1, N_j^2, ..., N_j^{t_j}$, where $t_j\geq 1$ and $N_j^1<N_j^2<...<N_j^{t_j}$. For a code of length $N_j^i$ and power $P_j$, we refer to it as an $(N_j^i, P_j)$-code in this paper. The probability that a user chooses to use the $(N_j^i, P_j)$-code is carefully designed such that, under the condition that all codewords with power higher than $P_j$ and those with the same power $P_j$ but longer code-lengths have been successfully decoded, the receiver can decode all $(N_j^i, P_j)$ codewords in the TIN way. 
  
  To design such a probability distribution, we first estimate the required signal-to-noise ratio (SNR) for different code-lengths to achieve the target block error rate $\epsilon$. A lower bound of the required SNR can be estimated using the following finite-length rate formula \cite{Polyanskiy2010FL}:
  \begin{equation}
  R \approx C(P)-\sqrt{\frac{V(P)}{N}}Q^{-1}(\epsilon),
  \end{equation}
  where $C(P)$ is the channel capacity and $V(P)$ is the channel dispersion. For a specific code (polar code in this paper), this can be done by simulations. After that, we calculate the number of users that can be supported by the $(N_j^i, P_j)$-code, denoted by $k_j^i$. The order of $k_j^i$ to be determined is:
  $$k_1^1\rightarrow k_1^2\rightarrow ... \rightarrow k_1^{t_1}\rightarrow k_2^1\rightarrow... k_s^{t_s-1}\rightarrow k_s^{t_s}.$$ As an example, suppose we are now to calculate $k_j^i$. 
  The average interference power to be treated as noise by users using the $(N_j^i, P_j)$-code is
  \begin{align}
  P_{int}=\sum_{a=1}^{j-1}\sum_{b=1}^{t_j}\frac{k_a^b N_a^b P_a}{N}+\sum_{b=1}^{i-1}\frac{k_j^b N_a^b P_j}{N}+ \frac{(k_j^i-1) N_j^i P_j}{N},
  \end{align}
  where the first item corresponds to interference caused by codewords with lower power, the second one corresponds to that caused by codewords with the same power but shorter code-length, and the last item is the interference from users adopting the same code. For a required SNR $snr$, the following inequality must hold to achieve the desired error rate:
  \begin{equation}
    \frac{P_j}{1+P_{int}}\geq snr.
  \end{equation}
  From this inequality we can calculate the maximum number of users $\bar{k}_j^i$ that can be supported by the $(N_j^i, P_j)$-code. To increase the robustness of the scheme, we assign $k_j^i$ to be smaller than $\bar{k}_j^i$ so that even if the number of users that select this code is more than expected in a certain range, their codewords are still decodable. 
  
  To initiate the calculation, we need to choose an $N_0\in\{N_1, N_2, ..., N_t\}$ as the shortest code-length to be used with the lowest power $P_1$ and the number of users $k_0$ that we wish this code to support. Suppose $snr_0$ is the required SNR for a length-$N_0$ code to achieve the target block error probability. Since the signal-to-interference-plus-noise ratio (SINR) is $\frac{P_1}{1+\frac{(k_0-1)N_0 P_1}{N}}$, the minimum required $P_1$ can be calculated as
  \begin{align}
  P_1=\frac{1}{\frac{1}{snr_0}-\frac{(k_0-1)N_0}{N}}.
  \end{align}
  As for $P_2,...,P_s$, their value can be flexibly chosen depending on the value of $K_a$ in order to achieve lower $E_b/N_0$.
  
  It should be noted that the number of power levels, the value of each power level, the number of code-lengths in each power level, the range of code-lengths and the designed user number of each code significantly affect system performance and should be carefully designed and optimized. Having determined all $k_j^i$, the probability that an active user chooses the $(N_j^i, P_j)$-code is set to be $$p_j^i=\frac{k_j^i}{K_a}.$$
  
  The code configuration information of each user, including code-length, power and interleaver, is represented by a $B_p$-bit preamble message and transmitted to the receiver by means of compressive sensing. For details of this part, we refer readers to \cite{pradhan2019joint}.
  
  \subsection{Decoding}
  \label{S:decoding}
  
  The preambles of active users are decoded by a compressive sensing decoder. To reduce the risk of outputting the wrong list of preambles, we can let the compressive sensing decoder output $K_b>K_a$ indices as suggested in \cite{pradhan2019joint}, since for the non-existing preambles, it is equivalent that an all-zero sequence is transmitted. We can also further apply a maximum likelihood decoding within the $K_b$ indices and output $K_a$ most likely ones, as suggested in \cite{Vem2017RA}.
  
  Having recovered the preambles, the receiver decodes in a TIN-SIC way. Note that the different interleavers used by different users help the receiver to distinguish each codeword's position. The decoding order of different codes is converse to the design order, i.e., codewords from the $(N_s^{t_s}, P_s)$ code are first decoded and subtracted from the received signal. Then the $(N_s^{t_s-1}, P_s)$ codewords are similarly decoded. After all codewords with power $P_s$ are decoded and subtracted, the receiver then decodes codewords with power $P_{s-1}$, and so on and so forth until all codewords are decoded. 
  
  \subsection{Complexity}
  
  Since we treat interference as noise in each step of decoding, any conventional polar decoding algorithms, such as CA-SCL decoding, can be used. For CA-SCL decoding, the complexity is $O(LN\log N)$, where $N$ is the code-length and $L$ is the decoding list size. Except there are extra zero-padding and interleaving procedures, the encoding and code construction complexity of our scheme is also similar to that of conventional polar coding, i.e., $O(N\log N)$ for encoding and $O(N)$ for code construction. Therefore, the per-user complexity of our scheme is just $O(LN\log N)$.
  
  \section{Numerical Results}
  \label{S:NResult}
  
  \subsection{TIN Performance of Polar Codes}
  Firstly, we determine the required SNR for polar codes of different lengths to achieve target block error rate $P_e\leq 0.05$ by simulations. The number of information bits to be transmitted is 85, and the code-lengths range from 768 to 8192. For code-lengths not equal to the power of 2, we use puncturing to shorten a longer polar code. Specifically, for a polar code of length $2^{m-1}<N<2^m$, we puncture the first $2^m-N$ bits from length-$2^m$ polar codes. We use the CA-SCL decoder with a large list size (L=8192) to see the potential of our scheme. Since the information payload is small, the CRC does not have to be very long. We find that CRC-12 can provide good performance. The result is shown in Table. \ref{example:polarperf}. 
  \begin{table}[htbp]
  	\centering
  	\caption{Required SNR for polar codes to achieve $P_e\leq 0.05$.}
  	\label{example:polarperf}
  	\begin{tabular}{|c|c|c|c|c|c|c|}
  		\hline
  		Code-length & 8192 & 7680 & 7168 & 6656 & 6144 & 5632 \\
  		\hline  		
  		Required SNR (dB)& -16.8 & -16.5 & -16.2 & -15.9 & -15.6 & -15.2 \\
  		\hline
  		Code-length & 5120 & 4608 & 4096 & 3584 & 3072 & 2560 \\
  		\hline  		
  		Required SNR (dB)& -14.7 & -14.2 & -13.9 & -13.4 & -12.6 & -11.8 \\
  		\hline
  		Code-length & 2048 & 1792 & 1536 & 1280 & 1024 & 768 \\
  		\hline  		
  		Required SNR (dB)& -10.9 & -10.3 & -9.6 & -8.7 & -7.8 & -6.5 \\
  		\hline
  	\end{tabular}
  \end{table}

\begin{figure}[tb]
  	\centering
  	\includegraphics[width=9cm]{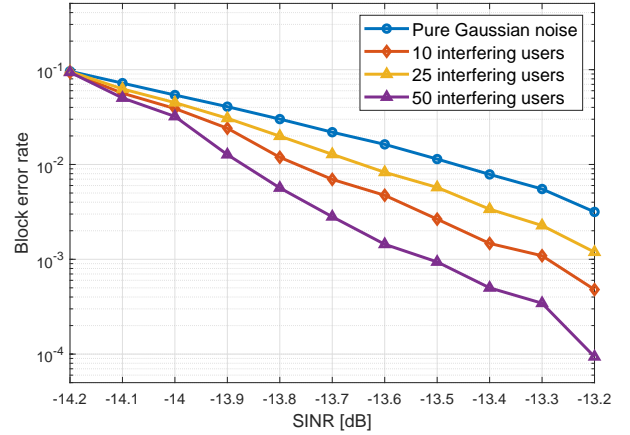}
  	\caption{Performance comparison of polar codes with the true and approximated interference.} \label{fig:perfComp}
  \end{figure}

  The simulations are performed using Gaussian noise. One may wonder if our approximation of interference degrades the actual performance. For this consideration, we have also run simulations to compare the performance of polar codes under the true interference introduced by our scheme and that under the approximated Gaussian interference. An example is shown in Fig. \ref{fig:perfComp}, in which we compare the performance of polar codes under various interfering users. Four cases are considered, i.e., pure Gaussian noise, Gaussian noise + 10 interfering users, Gaussian noise + 25 interfering users, and Gaussian noise + 50 interfering users. We normalize the BPSK symbol power to 1, and thus under the same SINR the noise variance in the four cases are different. A length-4096 polar code with 85 information bits and 12 CRC bits is used. The decoding is CA-SCL with $L=8192$. Interfering users also use length-4096 codes, and the interference is generated according to our sparse spreading scheme. For all four cases, decoding is performed under the Gaussian noise assumption. It can be seen that the approximation does not deteriorate error performance. On the contrary, the performance even slowly improves as the number of interfering user increases. It may look strange at first glance. This can partly be explained by that although the interference can be well approximated by a Gaussian noise, it is still discrete in nature. Thus, the uncertainty caused by the interference is in fact smaller than that caused by its Gaussian approximation. This result shows that using the estimated required SNRs in Table. \ref{example:polarperf} to estimate the performance of our scheme is reasonable (even a little conservative).

  \subsection{An example of unsourced Gaussian RAC}
  
  Now we consider a standard setup of $B=100$ information bits for each active user, $N=30000$ channel uses in total, target error probability of $\epsilon=0.05$, and various active user numbers ($K= 25 \sim 300$), which have been used in almost all previous schemes. Among the $B$ information bits, $B_p=15$ are used as preamble and encoded by a compressive sensing encoder of length $N_p=2000$ with power $P_p$, same as in \cite{pradhan2019joint}. The rest $B_c=85$ information bits are encoded using our proposed scheme and transmitted in the rest $N_c=28000$ channel uses. For simplicity, we set $P_j=2P_{j-1}$ for $j\geq 2$.  Since we use various code-lengths and power, the average energy-per-bit of the system is 
  \begin{equation}
  E_b/N_0= \frac{K_a N_p P_p+\sum_{j=1}^s \sum_{i=1}^{t_j}  k_j^i N_j^i P_j}{2K_a B_p+2B_c \sum_{j=1}^s \sum_{i=1}^{t_j} k_j^i}.
  \end{equation}
  
  The codebook configuration has significant impact on the energy-per-bit of the system. By trial and comparison, we find that for $K_a\leq 150$, a single power level is enough to provide good performance. As $K_a$ increases from 25 to 150, the optimal code-length number (that we have found so far) increases from 2 to 10. For example, for $K_a=150$, we use polar codes of length $3584,4096,...,8192$. The best codebook configuration that we have found is shown in Table \ref{example:fraction}. For $175\leq K_a \leq 250$, two power levels give better performance than one power level. The codebook configuration for $K_a=200$ is shown in Table \ref{example:fraction1}. For $K_a>250$, three power levels becomes most energy efficient. 
  
  \begin{table}[htbp]
  	\centering
  	\caption{Codebook configuration for $K_a=150$.}
  	\label{example:fraction}
  	\begin{tabular}{|c|c|c|c|c|c|}
  		\hline
  		Code-length & 8192 & 7680 & 7168 & 6656 & 6144  \\
  		\hline  		
  		Fraction& 0.073 & 0.073 & 0.073 & 0.073 & 0.1 \\
  		\hline
  		Code-length & 5632 & 5120 & 4608 & 4096 & 3584 \\
  		\hline  		
  		Fraction& 0.12 & 0.113 & 0.073 & 0.12 & 0.18 \\
  		\hline
  	\end{tabular}
  \end{table}

  \begin{table}[htbp]
  	\centering
  	\caption{Codebook configuration for $K_a=200$.}
  	\label{example:fraction1}
  	\begin{tabular}{|c|c|c|c|c|c|}
  		\hline
  		\multicolumn{6}{|c|}{$P_1$}\\
  		\hline
  		Code-length & 8192 & 7680 & 7168 & 6656 & 6144  \\
  		\hline  		
  		Fraction& 0.055 & 0.055 & 0.055 & 0.055 & 0.075 \\
  		\hline
  		Code-length & 5632 & 5120 & 4608 & 4096 & 3584 \\
  		\hline  		
  		Fraction& 0.09 & 0.085 & 0.055 & 0.09 & 0.135 \\
  		\hline
  		\multicolumn{6}{|c|}{$P_2=2P_1$}\\
  		\hline
  		Code-length & 5632 & 5120 & 4608 & 4096 &  \\
  		\hline  		
  		Fraction& 0.09 & 0.09 & 0.05 & 0.02 &   \\
  		\hline  
  	\end{tabular}
  \end{table}

  \begin{figure}[tb]
  	\centering
  	\includegraphics[width=9cm]{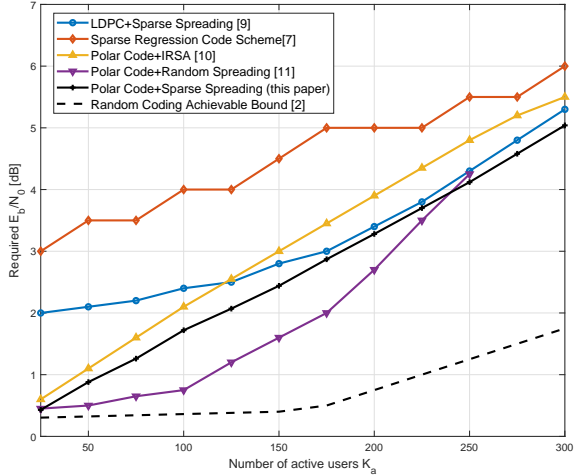}
  	\caption{Comparison of the required $E_b/N_0$ for various schemes.} \label{fig:EbN0Compare}
  \end{figure}
  
  For each $K_a$, we tried to find the optimal codebook configuration to minimize $E_b/N_0$, and the best result that we have found so far is shown in Fig. \ref{fig:EbN0Compare}. Note that reference \cite{pradhan2019polar} did not present the achievable $E_b/N_0$ for $K_a=275$ and 300. From the trend of their curve we can expect that it is worse than some other schemes. It can be observed that our scheme has the best performance for $K_a\geq 250$ and $K_a=25$, compared with four previous schemes. For $50\leq K_a \leq 225$, our scheme is only inferior to the random spreading scheme of \cite{pradhan2019polar}. Besides, it can be seen that $E_b/N_0$ (in dB) grows almost linearly with $K_a$ in our scheme, unlike the scheme of \cite{pradhan2019polar} which shows a drastic increase as $K_a$ goes large. In fact, we have also considered the cases of $K_a=325$ and $K_a=350$, and found that this linearity still holds.

 \section{Discussion}
 \label{S:Discussion}
 
 In this paper, we proposed a new low-complexity polar coding scheme for the unsourced Gaussian random access channel, based on sparse spreading and TIN-SIC. The proposed scheme demonstrates the best performance so far in the high user density region, and also provides competitive performance in the low and medium user density regions. Another advantage of our scheme is its low complexity, due to the fact that only single-user polar code is required. It would be interesting to study whether there are better codebook configurations that can achieve even better performance, since the example that we have provided may still be further improved.

\bibliographystyle{IEEEtran}
\bibliography{Polar_RA}

\begin{thebibliography}{10}
\providecommand{\url}[1]{#1}
\csname url@samestyle\endcsname
\providecommand{\newblock}{\relax}
\providecommand{\bibinfo}[2]{#2}
\providecommand{\BIBentrySTDinterwordspacing}{\spaceskip=0pt\relax}
\providecommand{\BIBentryALTinterwordstretchfactor}{4}
\providecommand{\BIBentryALTinterwordspacing}{\spaceskip=\fontdimen2\font plus
\BIBentryALTinterwordstretchfactor\fontdimen3\font minus
  \fontdimen4\font\relax}
\providecommand{\BIBforeignlanguage}[2]{{%
\expandafter\ifx\csname l@#1\endcsname\relax
\typeout{** WARNING: IEEEtran.bst: No hyphenation pattern has been}%
\typeout{** loaded for the language `#1'. Using the pattern for}%
\typeout{** the default language instead.}%
\else
\language=\csname l@#1\endcsname
\fi
#2}}
\providecommand{\BIBdecl}{\relax}
\BIBdecl

\bibitem{wu2020massive}
Y.~{Wu}, X.~{Gao}, S.~{Zhou}, W.~{Yang}, Y.~{Polyanskiy}, and G.~{Caire},
  ``Massive access for future wireless communication systems,'' \emph{IEEE
  Wireless Communications}, pp. 1--9, 2020.

\bibitem{Polyanskiy2017RA}
Y.~Polyanskiy, ``A perspective on massive random-access,'' in \emph{2017 IEEE
  International Symposium on Information Theory (ISIT)}, 2017, pp. 2523--2527.

\bibitem{Ordentlich2017RA}
O.~Ordentlich and Y.~Polyanskiy, ``Low complexity schemes for the random access
  {G}aussian channel,'' in \emph{2017 IEEE International Symposium on
  Information Theory (ISIT)}, 2017, pp. 2528--2532.

\bibitem{Vem2017RA}
A.~Vem, K.~R. Narayanan, J.~Cheng, and J.~Chamberland, ``A user-independent
  serial interference cancellation based coding scheme for the unsourced random
  access {G}aussian channel,'' in \emph{2017 IEEE Information Theory Workshop
  (ITW)}, 2017, pp. 121--125.

\bibitem{amalladinne2018coupled}
V.~K. Amalladinne, A.~Vem, D.~K. Soma, K.~R. Narayanan, and J.-F. Chamberland,
  ``A coupled compressive sensing scheme for unsourced multiple access,'' in
  \emph{2018 IEEE International Conference on Acoustics, Speech and Signal
  Processing (ICASSP)}.\hskip 1em plus 0.5em minus 0.4em\relax IEEE, 2018, pp.
  6628--6632.

\bibitem{amalladinne2018coded}
V.~K. Amalladinne, J.-F. Chamberland, and K.~R. Narayanan, ``A coded compressed
  sensing scheme for uncoordinated multiple access,'' \emph{arXiv preprint
  arXiv:1809.04745}, 2018.

\bibitem{fengler2019sparcs}
A.~Fengler, P.~Jung, and G.~Caire, ``{SPARC}s for unsourced random access,''
  \emph{arXiv preprint arXiv:1901.06234}, 2019.

\bibitem{calderbank2018chirrup}
R.~Calderbank and A.~Thompson, ``{CHIRRUP}: a practical algorithm for unsourced
  multiple access,'' \emph{arXiv preprint arXiv:1811.00879}, 2018.

\bibitem{pradhan2019joint}
A.~Pradhan, V.~Amalladinne, A.~Vem, K.~R. Narayanan, and J.-F. Chamberland, ``A
  joint graph based coding scheme for the unsourced random access {G}aussian
  channel,'' \emph{arXiv preprint arXiv:1906.05410}, 2019.

\bibitem{marshakov2019polar}
E.~Marshakov, G.~Balitskiy, K.~Andreev, and A.~Frolov, ``A polar code based
  unsourced random access for the {G}aussian {MAC},'' in \emph{2019 IEEE 90th
  Vehicular Technology Conference (VTC2019-Fall)}.\hskip 1em plus 0.5em minus
  0.4em\relax IEEE, 2019, pp. 1--5.

\bibitem{pradhan2019polar}
A.~K. Pradhan, V.~K. Amalladinne, K.~R. Narayanan, and J.-F. Chamberland,
  ``Polar coding and random spreading for unsourced multiple access,''
  \emph{arXiv preprint arXiv:1911.01009}, 2019.

\bibitem{arikan2009channel}
E.~Ar{\i}kan, ``Channel polarization: A method for constructing
  capacity-achieving codes for symmetric binary-input memoryless channels,''
  \emph{IEEE Transactions on Information Theory}, vol.~55, no.~7, pp.
  3051--3073, 2009.

\bibitem{tal2015list}
I.~{Tal} and A.~{Vardy}, ``List decoding of polar codes,'' \emph{IEEE
  Transactions on Information Theory}, vol.~61, no.~5, pp. 2213--2226, 2015.

\bibitem{Polyanskiy2010FL}
Y.~Polyanskiy, H.~V. Poor, and S.~Verd\'{u}, ``Channel coding rate in the
  finite blocklength regime,'' \emph{IEEE Transactions on Information Theory},
  vol.~56, no.~5, pp. 2307--2359, 2010.

\end{thebibliography}

\end{document}